\documentclass[sigconf]{acmart}
\usepackage{booktabs}

\AtBeginDocument{%
  }

\setcopyright{acmlicensed}
\copyrightyear{2025}
\acmYear{2025}

\acmConference[EARL Workshop@RecSys]{RecSys}{Sept 22--26,
  2025}{Prague, Czech Republic}

\begin{document}

\title{AudioBoost: Increasing Audiobook Retrievability in Spotify Search with Synthetic Query Generation}

\author{Enrico Palumbo$^1$, Gustavo Penha$^2$, Alva Liu$^3$, Marcus Eltscheminov$^4$, Jefferson Carvalho dos Santos$^4$, Alice Wang$^5$, Hugues Bouchard$^7$, Humberto Jesús Corona Pampin$^2$, Michelle Tran Luu$^6$}
\affiliation{
\institution{Spotify}
\country{$^1$Italy, $^2$Netherlands, $^3$Sweden, $^4$Brazil, $^5$USA, $^6$UK, $^7$Spain}
}
\email{{enricop,gustavop}@spotify.com}  


\renewcommand{\shortauthors}{Palumbo et al.}

\begin{abstract}
Spotify has recently introduced audiobooks as part of its catalog, complementing its music and podcast offering. Search is often the first entry point for users to access new items, and an important goal for Spotify is to support users in the exploration of the audiobook catalog. More specifically, we would like to enable users without a specific item in mind to broadly search by topic, genre, story tropes, decade, and discover audiobooks, authors and publishers they may like. To do this, we need to 1) inspire users to type more exploratory queries for audiobooks and 2) augment our retrieval systems to better deal with exploratory audiobook queries. This is challenging in a cold-start scenario, where we have a retrievabiliy bias due to the little amount of user interactions with audiobooks compared to previously available items such as music and podcast content. To address this, we propose AudioBoost, a system to boost audiobook retrievability in Spotify's Search via synthetic query generation. AudioBoost leverages Large Language Models (LLMs) to generate synthetic queries conditioned on audiobook metadata. The synthetic queries are indexed both in the Query AutoComplete (QAC) and in the Search Retrieval engine to improve query formulation and retrieval at the same time. We show through offline evaluation that synthetic queries increase retrievability and are of high quality. Moreover, results from an online A/B test show that AudioBoost leads to a +0.7\% in audiobook impressions, +1.22\% in audiobook clicks, and +1.82\% in audiobook exploratory query completions. 
\end{abstract}



\keywords{Synthetic Query Generation, Cold Start, Retrieval, Query Suggestion, LLMs}


\maketitle

\section{Introduction}

\begin{figure}[ht!]
    \centering
    \includegraphics[width=.5\textwidth, height=.24\textwidth]{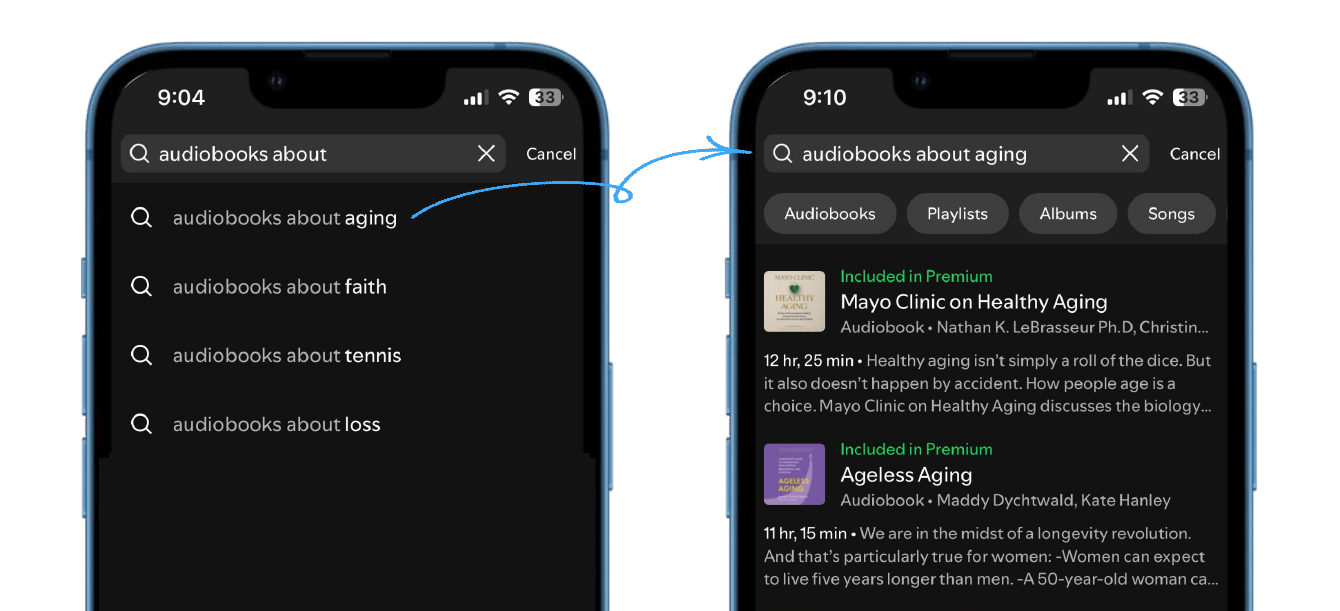}
    \caption{Illustrative example - Synthetic queries generated by an LLM are used as query completions to support query formulation (left) and for retrieval (right) to improve the retrievability of cold-start entities, i.e. audiobooks, in Spotify search.}
    \label{fig:quac_retrieval}
\end{figure}

Search engines in online content platforms have to disambiguate user queries against a huge number of potential catalog items across several possible content types. For instance, if the user searches for ``\textit{motivation}'' at Spotify, they could be looking for a specific song with the word \textit{motivation} in the title, a motivational playlist that fits their current vibe, or a podcast/audiobook about finding good habits that lead to motivation. To find the most relevant items for a user, modern search engines rely on a variety of signals, such as the user's history and the item's popularity. However, the estimation of the ``true'' item popularity is challenging in a cold-start scenario where a new type of content has been recently introduced in the platform. In such a scenario, most listeners are not used to engaging with the new content type, and interactions are mostly concentrated on previously available items, hence negatively impacting the retrievability (i.e. the chance of being retrieved) \cite{azzopardi2008retrievability} of new items. An important goal is then to raise awareness about the newly introduced items, helping users discover the breadth and depth of the catalog, while, at the same time, supporting authors and publishers in getting their work exposed and connecting with more audience.

Research on search mindsets \cite{searchmindsets} has highlighted that when users search for broader topics such as ``\textit{indie rock}'' or ``\textit{psychology}'' they are more prone to engage with system recommendations and suggestions, thus providing a chance to increase the retrievability of under-served content \cite{queryunderstandingunderserved}. Previous work has shown that query suggestions can be used effectively to lead users to type more exploratory, broad queries \cite{graphlearningexploratory, quac_paper}. In \cite{ctrlqgen} the authors propose a particularly promising paradigm to increase the retrievability of a set of target items that leverages LLMs to perform query generation controlling for the intent (i.e. broad vs narrow). More specifically, in \cite{ctrlqgen} the authors show that exploratory synthetic queries can be generated and used as 1) query suggestions influencing the query distribution toward more exploratory queries 2) to improve retrieval for exploratory queries via document augmentation. Crucially, both steps need to be done to maximize the effectiveness of the approach. While promising, so far this approach has been tested on general tail items rather than on new item types and has never been shown to work at scale in a production system. 

In this work, we describe AudioBoost, a system that increases audiobook retrievability in Spotify Search via synthetic query generation. AudioBoost leverages LLMs to generate synthetic queries conditioned on audiobook metadata (e.g. title, author, description). The LLM is prompted using a taxonomy that is defined for the audiobook use case and leveraging a chain-of-thought prompting strategy (Sec. \ref{sec:query_gen}). Then, the synthetic queries are used as candidates for query completions so that the users can be inspired to type more audiobook exploratory queries (Sec. \ref{sec:query_completion}). Finally, we perform document augmentation adding the synthetic queries to the audiobook representations in the retrieval system (Sec. \ref{sec:retrieval}). We use AudioBoost to generate synthetic queries for all audiobooks in the catalog and perform a set of offline evaluations to check the validity of the proposed approach in a controlled setting (Sec. \ref{sec:offline_results}). More in detail, we perform a simulation showing that the synthetic queries increase audiobook retrievability as expected (Sec. \ref{sec:retrievability_offline}), that the quality of the synthetic queries is high using an \emph{LLM-as-a-judge approach} (Sec. \ref{sec:llm_eval}). 
We finally test our method online in a large-scale A/B test, showing that it leads to +0.7\% in audiobook impressions, +1.22\% in audiobook clicks, and +1.82\% in audiobook exploratory query completions. 

In addition to being effective, AudioBoost is also appealing for practical reasons in an industry scenario. AudioBoost is highly scalable - the query generation and indexing steps are performed in offline pipelines, without affecting the latency and with a reasonable cost.

\section{Approach}
\begin{figure*}[ht!]
    \centering
    \includegraphics[width=.9\textwidth]{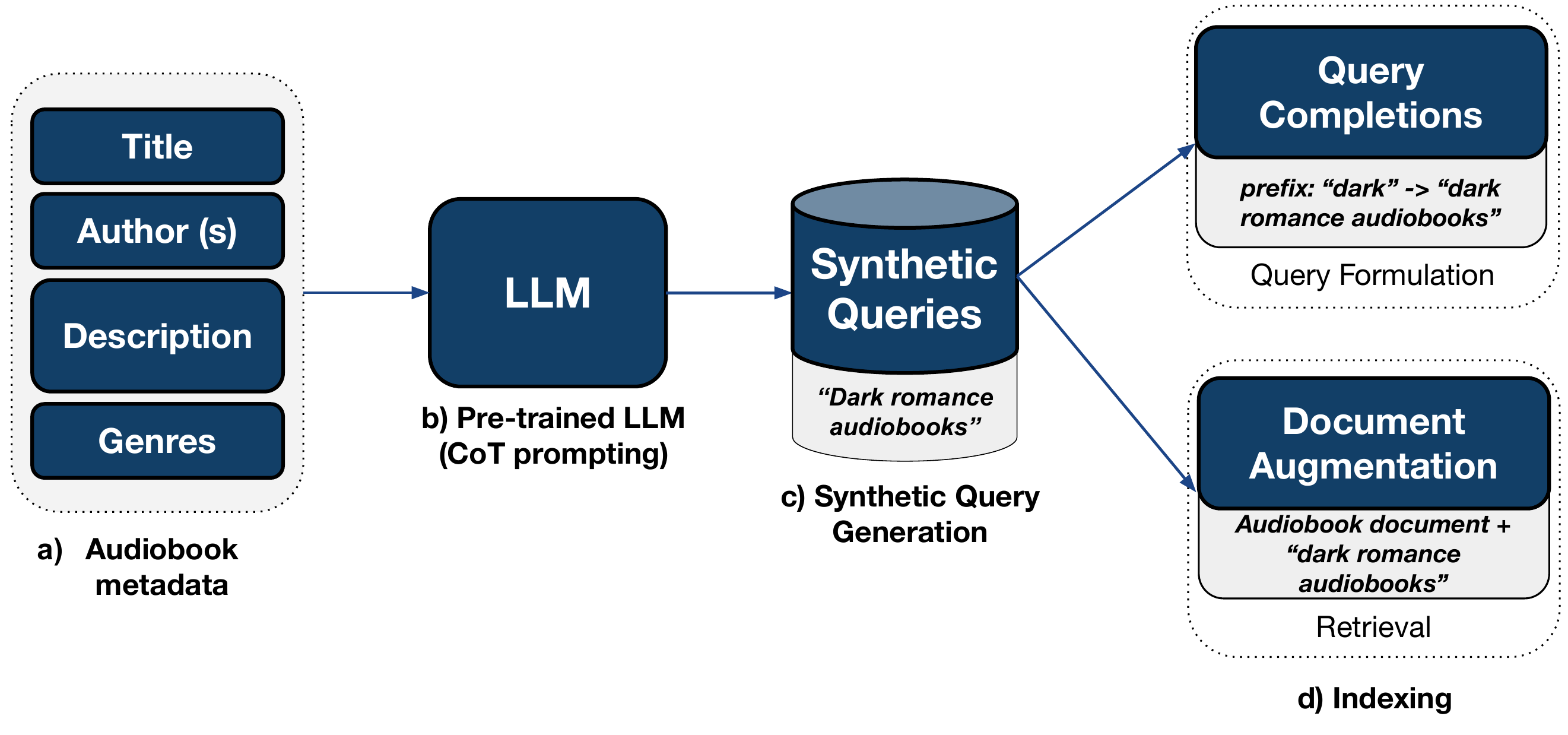}
    \caption{AudioBoost pipeline a) We collect metadata about a given audiobook b) we use a pre-trained LLM with chain-of-thought prompting to generate audiobook descriptors and synthetic queries c) we store the generated synthetic queries in a table d) we index synthetic queries as query completion and concatenate them to the document metadata in the retrieval system before indexing (document augmentation).}
    \label{fig:audioboost}
\end{figure*}

In this section, we describe how the AudioBoost pipeline works and its main components (Fig. \ref{fig:audioboost}).

\subsection{Query Generation}
\label{sec:query_gen}
To generate synthetic queries for audiobooks we first mapped out a taxonomy containing different types of audiobook descriptors. To define such taxonomy, we looked into how users search for audiobooks using both internal search queries and also requests issued at Reddit on the \textit{/r/booksuggestions/}\footnote{\url{https://www.reddit.com/r/booksuggestions/}} forum. For example, users might start a thread asking other users for ``\textit{books with a teen protagonist who overcomes many challenges}'', revealing that descriptions of the characters might serve as good queries. This manual approach has led to the following taxonomy: 

     \begin{enumerate}
          \item Genres, e.g. ``\textit{juvenile fiction}''. Genre-specific descriptors are those where users are seeking book recommendations within a particular literary genre such as horror, romance, detective, fantasy, etc.
         \item  Themes or topics, e.g. ``\textit{global politics}''. Descriptors based on specific subjects themes or topics such as societal issues, self-improvement, education, etc.
        \item Characters Descriptions, e.g. ``\textit{heroic protagonist}''. Descriptors about the personality, development, relationships, or unique characteristics of the protagonists or other significant characters within the story, e.g. heroic protagonist, book with great villains, etc.
        \item Moods, e.g. ``\textit{adventurous}''. Descriptors related to evoking a specific emotional response or mood, for example, books that make you cry, light books with elements of humor, dark or grim books, etc.
        \item Settings, e.g. ``\textit{China's Cultural Revolution.}''. Descriptors that focus on the reader's desire to immerse themselves in a specific location, period, or mood through the book they read, for example, Halloween reads, Christmas books, books set in Venice, books with cozy spooky magical vibes, etc.
        \item Personal situations, e.g. ``\textit{dealing with loss.}''. Descriptors tailored to specific circumstances or challenges people are facing in their personal lives, e.g. parenting and family, dealing with loss or grief, books on relationships, personal life transition, etc.
        \item Story tropes, e.g. ``\textit{enemies to lovers.}''. Descriptors that explore specific narrative themes or plot devices commonly found in literature. These tropes provide a framework for the story and often resonate with readers due to their familiarity and emotional impact, e.g. coming-of-age, forbidden love, found family, characters don't initially like each other, starting from a rough spot, etc.
        \item Target audiences, e.g. ``\textit{children's literature.}''. Descriptors based on the target audience of the book, e.g. family time, bedtime stories for adults. 
        \item Objective-based, e.g. ``\textit{to learn Japanese.}''. descriptors related to a specific activity or objective, e.g. meditation, sleep stories, etc.
        \item Named entities, e.g. ``\textit{Britney Spears.}''. Non-fictional entities related to the book.
    \end{enumerate}

Based on this taxonomy we use a prompt approach to generate queries in a chain-of-though manner. Before asking the LLM to generate queries, we first ask it to generate descriptors under the 10 categories of this taxonomy and then to use such descriptors to generate two types of synthetic queries:

     \begin{enumerate}
         \item[(11)] Queries, e.g. ``\textit{realistic fiction audiobooks}''.
         \item[(12)] Compound queries, e.g. ``\textit{Stephen King supernatural fiction audiobooks}''.
    \end{enumerate}

\textit{Queries} use broader descriptors and combinations from the previous types with the ``\textit{audiobook}'' suffix, whereas \textit{Compound queries} leverage a combination of narrow attributes (author names) and broader descriptors. Besides instructions on how to generate descriptors for each of the 12 types described above, we also use two in-context learning examples in the prompt. 

The metadata used for each audiobook as input to the model is the audiobook title, audiobook author(s), audiobook description, and BISAC genres\footnote{\url{https://www.bisg.org/BISAC-Subject-Codes-main}}. For each audiobook in the catalog, we use an LLM to generate the 12 types of descriptors.

\subsection{Query Completions}
\label{sec:query_completion}
To inspire users to type more exploratory queries for audiobooks and support them at the query formulation stage, we index the synthetic queries in the Query AutoComplete system. The Query AutoComplete (QAC) system presents query completions as the user types a query, matching the current prefix. For instance, if a user is typing a prefix ``\emph{audiob}'' we want to be able to suggest different exploration ideas such as ``\emph{audiobooks for children}'', ``\emph{audiobooks about love}'', ``\emph{audiobooks in french}''.
The QAC system has multiple sources of query completions, including item titles, complete queries from the search logs, and synthetic queries. For a given query prefix $p$, each source $j$ produces a set of $K$ candidate query completions $q_{jk}$ with $k=1, .., K$. The set of $K$ candidate query completions per source is determined by a predefined prefix matching score $v = v(p, q_{jk})$ and a global score $s = s(q_{jk})$ that models the inherent likelihood of the query completion $q_{jk}$. Then, all candidate query completions are combined and a re-ranker model is trained to select the top-N completions that are most relevant for the user prefix based on a number of features such as prefix matching, popularity, and user preferences \cite{quac_paper}.

We create a new source of query completions with synthetic queries for audiobooks. To define the global score $s$ we cannot rely on historical data for the popularity of the query, as most synthetic queries are not well represented in the search logs. Hence, we rely on the popularity of the audiobooks associated with the queries instead, compounded with a score that measures the broadness of the query, since we aim to prioritize broad queries that favor the catalog exploration. More specifically, given a query $q$ such as ``\emph{ancient history audiobooks}'' and a list of audiobooks $A = \{a_i\}$ from which $q$ was generated, we define:
\begin{equation}
    s = median\_popularity(q) * broadness(q)
\end{equation}
where $broadness(q) = log(|A| + 1)$. Broadness is higher for broad synthetic queries that are associated with many distinct audiobooks such as ``\emph{ancient history audiobooks}'' and lower for highly specific queries such as ``\emph{jrr tolkien first audiobooks}''. 

\subsection{Document Augmentation}
\label{sec:retrieval}
To improve the number of queries that retrieve audiobooks for such synthetic broad queries we rely here on document expansion for sparse retrieval systems. Similar to \textit{doc2query}~\cite{nogueira2019document,nogueira2019doc2query}, besides indexing the existing metadata of each audiobook, we also index the synthetic descriptors generated for them (types 1--12 from Sec~\ref{sec:query_gen}):

\begin{itemize}
  \item document = "title - author - description - genres"
  \item augmented\_document = "title - author - description - genres - descriptors - synthetic queries"
\end{itemize}

The sparse retrieval system is based on BM25 ~\cite{robertson2009probabilistic}, a keyword-matching system that models documents as a bag of words and assigns weights based on word frequencies. In this context, doing document augmentation with the synthetic queries has two potential effects: (1) modifying the weight of words that already exist in the item metadata by repeating them and (2) adding new words that were not covered by the audiobook metadata.

\section{Offline Results}
\label{sec:offline_results}
In this section, we describe the results of our offline experimentation where we tested the increase in retrievability due to the synthetic query generation in different conditions, the quality of synthetic queries as evaluated with an \textit{LLM-as-a-judge} approach.
\subsection{Retrievability Simulation}
\label{sec:retrievability_offline}
\paragraph{Dataset} For this experiment, we used a sample of data containing search successes from user logs. Specifically, we took a sample of query and entity pairs from a single day for three entity types at the platform: audiobooks, playlists, and podcast shows, where the user country is \textit{US}, the query is a reformulation (not the initial query issued by the user in our instant search system) and with more than 5 characters. The resulting distribution of distinct queries to entity types is as follows: 12.77\% for audiobooks, 43.47\% for playlists, and 43.76\% for podcasts.

\paragraph{Evaluation methodology}
We employ a BM25 model (we use Pyterrier~\cite{macdonald2021pyterrier} implementation with default hyperparameters), indexing each entity with their textual metadata (title, description, and genres when available). We have four different configurations in this experiment. Configuration 1 is the baseline, where synthetic queries are not used. Configurations 2 and 3 are partial solutions that do only one part of the proposed solution at a time (document expansion with synthetic queries in configuration 2 and synthetic query suggestion in configuration 3). Configuration 4 uses the synthetic queries in both query suggestion and document expansion.


\paragraph{Evaluation metric}
The retrievability of an entity $e$, as defined by~\cite{azzopardi2008retrievability}: $r(\mathbf{e})=\sum_{\mathbf{q} \in \mathbf{Q}} o_{q} \cdot f\left(k_{e q}, c\right)$, where $\mathbf{Q}$ is the set of queries, $o_{q}$ is the weight of each query---here we use 1 for all queries---and $f\left(k_{e q}, c\right)$ is 1 if the entity $e$ is ranked above $c$ by the search system (in our experiments we set $c=100$) and 0 otherwise. For configurations 1 and 2, the set of queries is a sample of 20k queries from the logs. Adding a sample of the synthetic queries (also 20k queries) for the query set $\mathbf{Q}$ used to calculate the retrievability (configurations 3 and 4) in this simulation assumes that such synthetic query suggestions made by the system would be clicked by at least one user. The \textbf{retrievability share} of an entity type is the sum of the retrievability for entities of that type. The retrievability percentage share of each entity type is calculated by dividing the sum of the retrievability of the entities of that type by the sum of retrievability for all entities.

\paragraph{Results}
\begin{table}[ht!]
 \caption{Results for the retrievability simulation when all suggested synthetic queries are clicked.}
 \label{table:retrievability_results}
\begin{tabular}{@{}lll@{}}
\toprule
\multicolumn{3}{c}{Retrievability percentage share} \\ \midrule
↓Query set $\backslash$ Retrieval →  & BM25 & \begin{tabular}[c]{@{}l@{}}BM25 \\ \footnotesize{+ doc augmentation} \end{tabular} \\ \midrule
Queries from logs & {\begin{tabular}[c]{@{}l@{}}\textbf{Configuration 1}\\ \textit{audiobook}:   23.36 \%\\ \textit{playlist}:         17.84 \%\\ \textit{podcast}:          58.78 \%\end{tabular}} & {\begin{tabular}[c]{@{}l@{}}\textbf{Configuration 2}\\ \textit{audiobook}:   26.20 \%\\ \textit{playlist}:         16.67 \%\\ \textit{podcast}:          57.11 \%\end{tabular}} \\ \midrule
\begin{tabular}[c]{@{}l@{}}Queries from logs \\ \footnotesize{+ query completions} \end{tabular} & {\begin{tabular}[c]{@{}l@{}}\textbf{Configuration 3}\\ \textit{audiobook}:    34.49 \%\\ \textit{playlist}:            9.72 \%\\ \textit{podcast}:           55.78 \%\end{tabular}} & {\begin{tabular}[c]{@{}l@{}}\textbf{Configuration 4}\\ \textit{audiobook}:    \textbf{56.97} \%\\ \textit{playlist}:            8.13 \%\\ \textit{podcast}:           34.89 \%\end{tabular}} \\ \bottomrule
\end{tabular}
\end{table}
Table~\ref{table:retrievability_results} shows the results of this simulation, for each configuration considering that the entire set of synthetic queries suggested would receive clicks. In configuration 2, some queries from the logs that would retrieve playlists or shows now return more audiobooks. For example, the following queries from the logs ``\textit{audiobooks}'', ``\textit{growth}'' and ``\textit{christian}'' retrieve more audiobooks when compared to configuration 1, due to synthetic queries that were added to the representation of audiobooks (i.e. document expansion) such as ``\textit{Self-Help Audiobooks for Spiritual Growth}'' and ``\textit{Christian Audiobooks}''.

In configuration 3, we change the query distribution to have more audiobook-focused queries in the system (remember that the original distribution of the dataset is skewed towards queries for playlists and podcast shows), and thus we can increase the number of audiobooks retrieved, as some of the synthetic queries will match with the title, description, and genre metadata already available.

The most effective approach though is a combination of both helping users to issue more broad audiobook queries and also modifying the retrieval system so that such queries lead to audiobooks (configuration 4).

\begin{figure}[ht!]
    \centering
    \includegraphics[width=.4\textwidth]{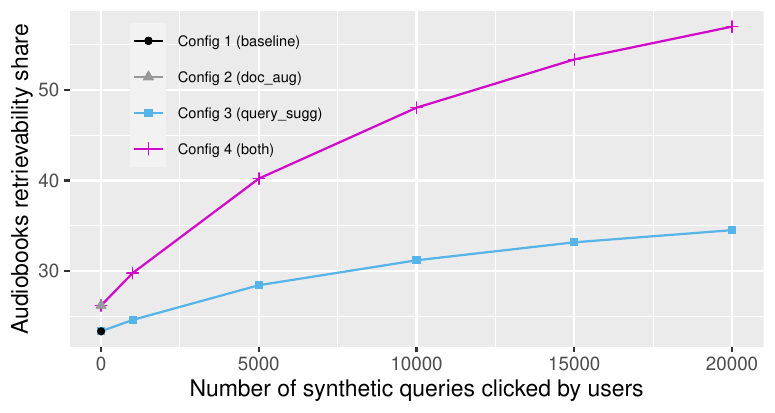}
    \caption{Offline simulation showing the impact on retrievability share of audiobooks when increasing the number of clicks towards suggested synthetic query completions.}
    \label{fig:retrievability_plot}
\end{figure}

We can see in Figure~\ref{fig:retrievability_plot} that the increase in retrievability share for audiobooks increases as long as more users click on the suggested queries when using the proposed approach (configuration 4).

\subsection{LLM evaluation}
\label{sec:llm_eval}
We evaluate the quality of the synthetic queries using an \emph{LLM-as-a-judge} approach \cite{zheng2023judging,rahmani2024llmjudge}, where a powerful LLM is used as an evaluator of another LLM's output. This approach has been shown to correlate with human evaluation both in public benchmarks \cite{zheng2023judging,rahmani2024llmjudge} and in internal assessments on similar tasks. 

We instruct the evaluator $LLM_{eval}$  using a few-shot prompting approach to judge synthetic queries on several dimensions:
\begin{itemize}
    \item \emph{quality}: queries need to be complete, well-formatted, without misspellings
    \item \emph{relevancy}: queries need to be relevant to the audiobook metadata that was provided as input
    \item \emph{diversity}: queries need to be not redundant (at the audiobook level). 
    \item \emph{broadness}: queries need to refer to general topics, and genres, rather than to specific audiobooks
\end{itemize}
All metrics are boolean and evaluated at the query level, except for \emph{diversity} which is a single value for the group of queries associated with an audiobook. We obtain high scores on all dimensions: $quality = 99.7\%$, $relevancy = 97.2\%$, $diversity = 81.5\%$, $broadness = 84.2\%$.

\section{Online Results}
We run a large-scale A/B test for 3 weeks comparing the default QAC and retrieval system to a treatment that uses AudioBoost, namely where we 1) add the synthetic queries as an additional source of query completions 2) use the synthetic queries to perform document augmentation in a sparse retrieval system (Fig. \ref{fig:audioboost}). 

We measure several metrics online to account for audiobook retrievability and exploratory searches for audiobooks at the SERP (Search Engine Results Page) level:
\begin{itemize}
    \item \emph{impressions}: number of audiobook impressions per SERP
    \item \emph{clicks}: number of clicks on audiobooks per SERP 
    \item \emph{coverage}: overall number of query completions shown per SERP
    \item \emph{exploration}: number of clicks on exploratory\footnote{The classification of exploratory queries is associated to the broadness of the query, i.e. number of distinct items they are associated with, and other factors.} query completions leading to audiobook interactions per SERP
\end{itemize}
We observe that AudioBoost leads to +0.7\% in impressions, +1.22\% in clicks, +0.03\% in coverage, and +1.82\% in exploration. At the same time, guardrail metrics that check for the overall engagement with QAC and overall search effectiveness are neutral. All reported results are statistically significant with a t-test with a p-value of 1\%.

\section{Conclusions}
In this work, we have introduced AudioBoost, a system to increase audiobook retrievability in Spotify search via synthetic query generation. 
Building up on previous research work on query generation, we have shown that synthetic query generation is a viable strategy to support exploratory search in a cold-start scenario where new item types have been introduced in the catalog in a production setting. We have used LLMs to generate synthetic queries for audiobooks and we have indexed them at the query formulation stage, inspiring users to type more exploratory queries for audiobooks, and at the retrieval stage, fulfilling such queries with relevant search results. Offline results have confirmed the validity of the proposed solution, in terms of retrievability, coverage increases and query quality. Online results show its effectiveness at scale in a real production system with millions of users. 

AudioBoost increases impressions and clicks on audiobooks, inspiring users to explore the catalog and helping authors and publishers gain visibility and engage with new listeners. More in general, this work highlights that synthetic query generation is a powerful strategy to increase the visibility of a set of target items in a search system. AudioBoost is also appealing from the productionalization point of view, as the query generation and indexing steps can be performed offline in a batch pipeline, with reasonable cost and without affecting the latency of the QAC and retrieval systems. Given these results, we have rolled out AudioBoost in production.

It is important to notice that, while this work focuses on the retrieval stage, many of these techniques could be applied similarly to the re-ranking stage of a typical search engine system, and in the future, we plan to investigate this direction further. 

We also aim to further explore the intersection of synthetic query generation with other approaches that have shown to work well to address the cold-start problem, such as explore-exploit strategies and/or content-based recommendations.

\section*{Acknowledgments}
We would like to thank Anders Nyman, Carlotta Balboni, Emelie Jin, Jonny Brooks-Bartlett, Yannis Kantarelis for the useful discussions and their help in iterating on the technology and driving the project forward.

\bibliographystyle{ACM-Reference-Format}
\bibliography{sample-base}


\begin{thebibliography}{12}


\ifx \showCODEN    \undefined \def \showCODEN     #1{\unskip}     \fi
\ifx \showDOI      \undefined \def \showDOI       #1{#1}\fi
\ifx \showISBNx    \undefined \def \showISBNx     #1{\unskip}     \fi
\ifx \showISBNxiii \undefined \def \showISBNxiii  #1{\unskip}     \fi
\ifx \showISSN     \undefined \def \showISSN      #1{\unskip}     \fi
\ifx \showLCCN     \undefined \def \showLCCN      #1{\unskip}     \fi
\ifx \shownote     \undefined \def \shownote      #1{#1}          \fi
\ifx \showarticletitle \undefined \def \showarticletitle #1{#1}   \fi
\ifx \showURL      \undefined \def \showURL       {\relax}        \fi
\providecommand\bibfield[2]{#2}
\providecommand\bibinfo[2]{#2}
\providecommand\natexlab[1]{#1}
\providecommand\showeprint[2][]{arXiv:#2}

\bibitem[Azzopardi and Vinay(2008)]%
        {azzopardi2008retrievability}
\bibfield{author}{\bibinfo{person}{Leif Azzopardi} {and} \bibinfo{person}{Vishwa Vinay}.} \bibinfo{year}{2008}\natexlab{}.
\newblock \showarticletitle{Retrievability: An evaluation measure for higher order information access tasks}. In \bibinfo{booktitle}{\emph{Proceedings of the 17th ACM conference on Information and knowledge management}}. \bibinfo{pages}{561--570}.
\newblock


\bibitem[Li et~al\mbox{.}(2019)]%
        {searchmindsets}
\bibfield{author}{\bibinfo{person}{Ang Li}, \bibinfo{person}{Jennifer Thom}, \bibinfo{person}{Praveen Chandar}, \bibinfo{person}{Christine Hosey}, \bibinfo{person}{Brian~St. Thomas}, {and} \bibinfo{person}{Jean Garcia-Gathright}.} \bibinfo{year}{2019}\natexlab{}.
\newblock \showarticletitle{Search Mindsets: Understanding Focused and Non-Focused Information Seeking in Music Search}. In \bibinfo{booktitle}{\emph{The World Wide Web Conference}} (San Francisco, CA, USA) \emph{(\bibinfo{series}{WWW '19})}. \bibinfo{publisher}{Association for Computing Machinery}, \bibinfo{address}{New York, NY, USA}, \bibinfo{pages}{2971–2977}.
\newblock
\showISBNx{9781450366748}
\urldef\tempurl%
\url{https://doi.org/10.1145/3308558.3313627}
\showDOI{\tempurl}


\bibitem[Lindstrom et~al\mbox{.}(2024)]%
        {quac_paper}
\bibfield{author}{\bibinfo{person}{Henrik Lindstrom}, \bibinfo{person}{Humberto~Jesus Corona~Pampin}, \bibinfo{person}{Enrico Palumbo}, {and} \bibinfo{person}{Alva Liu}.} \bibinfo{year}{2024}\natexlab{}.
\newblock \showarticletitle{Encouraging Exploration in Spotify Search through Query Recommendations}. In \bibinfo{booktitle}{\emph{Proceedings of the 18th ACM Conference on Recommender Systems}} (Bari, Italy) \emph{(\bibinfo{series}{RecSys '24})}. \bibinfo{publisher}{Association for Computing Machinery}, \bibinfo{address}{New York, NY, USA}, \bibinfo{pages}{775–777}.
\newblock
\showISBNx{9798400705052}
\urldef\tempurl%
\url{https://doi.org/10.1145/3640457.3688035}
\showDOI{\tempurl}


\bibitem[Macdonald et~al\mbox{.}(2021)]%
        {macdonald2021pyterrier}
\bibfield{author}{\bibinfo{person}{Craig Macdonald}, \bibinfo{person}{Nicola Tonellotto}, \bibinfo{person}{Sean MacAvaney}, {and} \bibinfo{person}{Iadh Ounis}.} \bibinfo{year}{2021}\natexlab{}.
\newblock \showarticletitle{PyTerrier: Declarative experimentation in Python from BM25 to dense retrieval}. In \bibinfo{booktitle}{\emph{Proceedings of the 30th acm international conference on information \& knowledge management}}. \bibinfo{pages}{4526--4533}.
\newblock


\bibitem[Nogueira et~al\mbox{.}(2019a)]%
        {nogueira2019doc2query}
\bibfield{author}{\bibinfo{person}{Rodrigo Nogueira}, \bibinfo{person}{Jimmy Lin}, {and} \bibinfo{person}{AI Epistemic}.} \bibinfo{year}{2019}\natexlab{a}.
\newblock \showarticletitle{From doc2query to docTTTTTquery}.
\newblock \bibinfo{journal}{\emph{Online preprint}}  \bibinfo{volume}{6} (\bibinfo{year}{2019}).
\newblock


\bibitem[Nogueira et~al\mbox{.}(2019b)]%
        {nogueira2019document}
\bibfield{author}{\bibinfo{person}{Rodrigo Nogueira}, \bibinfo{person}{Wei Yang}, \bibinfo{person}{Jimmy Lin}, {and} \bibinfo{person}{Kyunghyun Cho}.} \bibinfo{year}{2019}\natexlab{b}.
\newblock \showarticletitle{Document expansion by query prediction}.
\newblock \bibinfo{journal}{\emph{arXiv preprint arXiv:1904.08375}} (\bibinfo{year}{2019}).
\newblock


\bibitem[Palumbo et~al\mbox{.}(2023)]%
        {graphlearningexploratory}
\bibfield{author}{\bibinfo{person}{Enrico Palumbo}, \bibinfo{person}{Andreas Damianou}, \bibinfo{person}{Alice Wang}, \bibinfo{person}{Alva Liu}, \bibinfo{person}{Ghazal Fazelnia}, \bibinfo{person}{Francesco Fabbri}, \bibinfo{person}{Rui Ferreira}, \bibinfo{person}{Fabrizio Silvestri}, \bibinfo{person}{Hugues Bouchard}, \bibinfo{person}{Claudia Hauff}, \bibinfo{person}{Mounia Lalmas}, \bibinfo{person}{Ben Carterette}, \bibinfo{person}{Praveen Chandar}, {and} \bibinfo{person}{David Nyhan}.} \bibinfo{year}{2023}\natexlab{}.
\newblock \showarticletitle{Graph Learning for Exploratory Query Suggestions in an Instant Search System}. In \bibinfo{booktitle}{\emph{Proceedings of the 32nd ACM International Conference on Information and Knowledge Management}} (Birmingham, United Kingdom) \emph{(\bibinfo{series}{CIKM '23})}. \bibinfo{publisher}{Association for Computing Machinery}, \bibinfo{address}{New York, NY, USA}, \bibinfo{pages}{4780–4786}.
\newblock
\showISBNx{9798400701245}
\urldef\tempurl%
\url{https://doi.org/10.1145/3583780.3615481}
\showDOI{\tempurl}


\bibitem[Penha et~al\mbox{.}(2023)]%
        {ctrlqgen}
\bibfield{author}{\bibinfo{person}{Gustavo Penha}, \bibinfo{person}{Enrico Palumbo}, \bibinfo{person}{Maryam Aziz}, \bibinfo{person}{Alice Wang}, {and} \bibinfo{person}{Hugues Bouchard}.} \bibinfo{year}{2023}\natexlab{}.
\newblock \showarticletitle{Improving Content Retrievability in Search with Controllable Query Generation}. In \bibinfo{booktitle}{\emph{Proceedings of the ACM Web Conference 2023}} (Austin, TX, USA) \emph{(\bibinfo{series}{WWW '23})}. \bibinfo{publisher}{Association for Computing Machinery}, \bibinfo{address}{New York, NY, USA}, \bibinfo{pages}{3182–3192}.
\newblock
\showISBNx{9781450394161}
\urldef\tempurl%
\url{https://doi.org/10.1145/3543507.3583261}
\showDOI{\tempurl}


\bibitem[Rahmani et~al\mbox{.}(2024)]%
        {rahmani2024llmjudge}
\bibfield{author}{\bibinfo{person}{Hossein~A Rahmani}, \bibinfo{person}{Emine Yilmaz}, \bibinfo{person}{Nick Craswell}, \bibinfo{person}{Bhaskar Mitra}, \bibinfo{person}{Paul Thomas}, \bibinfo{person}{Charles~LA Clarke}, \bibinfo{person}{Mohammad Aliannejadi}, \bibinfo{person}{Clemencia Siro}, {and} \bibinfo{person}{Guglielmo Faggioli}.} \bibinfo{year}{2024}\natexlab{}.
\newblock \showarticletitle{LLMJudge: LLMs for Relevance Judgments}.
\newblock \bibinfo{journal}{\emph{arXiv preprint arXiv:2408.08896}} (\bibinfo{year}{2024}).
\newblock


\bibitem[Robertson et~al\mbox{.}(2009)]%
        {robertson2009probabilistic}
\bibfield{author}{\bibinfo{person}{Stephen Robertson}, \bibinfo{person}{Hugo Zaragoza}, {et~al\mbox{.}}} \bibinfo{year}{2009}\natexlab{}.
\newblock \showarticletitle{The probabilistic relevance framework: BM25 and beyond}.
\newblock \bibinfo{journal}{\emph{Foundations and Trends{\textregistered} in Information Retrieval}} \bibinfo{volume}{3}, \bibinfo{number}{4} (\bibinfo{year}{2009}), \bibinfo{pages}{333--389}.
\newblock


\bibitem[Tomasi et~al\mbox{.}(2020)]%
        {queryunderstandingunderserved}
\bibfield{author}{\bibinfo{person}{Federico Tomasi}, \bibinfo{person}{Rishabh Mehrotra}, \bibinfo{person}{Aasish Pappu}, \bibinfo{person}{Judith B\"{u}tepage}, \bibinfo{person}{Brian Brost}, \bibinfo{person}{Hugo Galv\~{a}o}, {and} \bibinfo{person}{Mounia Lalmas}.} \bibinfo{year}{2020}\natexlab{}.
\newblock \showarticletitle{Query Understanding for Surfacing Under-served Music Content}. In \bibinfo{booktitle}{\emph{Proceedings of the 29th ACM International Conference on Information \& Knowledge Management}} (Virtual Event, Ireland) \emph{(\bibinfo{series}{CIKM '20})}. \bibinfo{publisher}{Association for Computing Machinery}, \bibinfo{address}{New York, NY, USA}, \bibinfo{pages}{2765–2772}.
\newblock
\showISBNx{9781450368599}
\urldef\tempurl%
\url{https://doi.org/10.1145/3340531.3412741}
\showDOI{\tempurl}


\bibitem[Zheng et~al\mbox{.}(2023)]%
        {zheng2023judging}
\bibfield{author}{\bibinfo{person}{Lianmin Zheng}, \bibinfo{person}{Wei-Lin Chiang}, \bibinfo{person}{Ying Sheng}, \bibinfo{person}{Siyuan Zhuang}, \bibinfo{person}{Zhanghao Wu}, \bibinfo{person}{Yonghao Zhuang}, \bibinfo{person}{Zi Lin}, \bibinfo{person}{Zhuohan Li}, \bibinfo{person}{Dacheng Li}, \bibinfo{person}{Eric Xing}, {et~al\mbox{.}}} \bibinfo{year}{2023}\natexlab{}.
\newblock \showarticletitle{Judging llm-as-a-judge with mt-bench and chatbot arena}.
\newblock \bibinfo{journal}{\emph{Advances in Neural Information Processing Systems}}  \bibinfo{volume}{36} (\bibinfo{year}{2023}), \bibinfo{pages}{46595--46623}.
\newblock


\end{thebibliography}

\end{document}